\documentclass[twocolumn,showpacs,preprintnumbers,amsmath,amssymb]{revtex4}
\usepackage{graphicx}% Include figure files
\usepackage{dcolumn}% Align table columns on decimal point
\usepackage{bm}% bold math

\begin{document}

\preprint{}

\title{Transport and the Order Parameter of Superconducting UPt$_{3}$}

\author{W. C. Wu$^1$ and R. Joynt$^2$}
%\altaffiliation[Also at ]{Physics Department, XYZ University.}
\affiliation{%
$^1$Department of Physics, National Taiwan Normal
University, Taipei 11650, Taiwan\\
$^2$Department of Physics,
University of Wisconsin-Madison, Madison, WI 53706
}%

\date{\today}

\begin{abstract}
We calculate the ultrasonic absorption and the thermal
conductivity in the superconducting state of UPt$_{3}$ as
functions of temperature and direction of propagation and
polarization. Two leading candidates for the superconducting order
parameter are considered: the $E_{1g}$ and $E_{2u}$
representations. Both can fit the data except for the ultrasonic
absorption in the $A$ phase. To do that, it is necessary to
suppose that the system has only a single domain, and that must be
chosen as the most favorable one. However, the $E_{2u}$ theory
requires fine-tuning of parameters to fit the low temperature
thermal conductivity. Thus, transport data favor the $E_{1g}$
theory.  Measurements of the thermal conductivity as a function of
pressure at low temperature could help to further distinguish the
two theories.
\end{abstract}

\maketitle

\section{\label{sec:intro}Introduction}

UPt$_{3}$ continues to pose a fundamental challenge to solid-state
physics. It becomes superconducting at about $T_{c+}=0.5K$, entering the $%
A $ phase, and has a second transition at about $T_{c-}=0.44K$ entering the $%
B$ phase. Its unusual phase diagram and thermodynamics show that
it is unconventional. The precise nature of its superconducting
order parameter remains elusive, however.

From the purely theoretical point of view, order parameters (OPs)
that transform according to a single representation of the
$D_{6h}$ group and that take into account spin-orbit coupling are
the most desirable. This includes the $E_{1g}$ \cite{joynt1} and
$E_{2u}$ \cite{sauls1} theories. Because of the perception that
these theories may not give sufficient phenomenological
flexibility, other pictures have been constructed. One of these is
the mixed-representation ({\it e.g.}, $A$-$B$) \cite{garg}
picture. This suffers from the problem of fine tuning, since the
coupling constants of the two representations must be nearly
identical to account for the nearness of $T_{c+}$ and $T_{c-}$.
Another picture is the extended-symmetry-group spin triplet theory
\cite{machida}. However, this requires zero spin-orbit coupling in
an $f$-electron system, an artificial assumption.

From the point of view of comparison to experiment, the most
stringent constraints on theory come from detailed measurements of
the thermal conductivity $\kappa$ \cite{thermalc} and the
ultrasonic attenuation $\alpha $ \cite{alpha}. Because of the
directional information inherent in these probes, they offer the
best chance to distinguish between order parameters that may have
only relatively subtle differences \cite{moreno96}. This
expectation is reinforced by the fact that the ultrasonic
attenuation has been measured in the $A$ phase, the most
anisotropic of the zero-field phases. Previous calculations have
indicated that the mixed representation and extended symmetry
group scenarios are probably inconsistent with these measurements
\cite{sauls2}. Accordingly we focus in this paper on the $E_{1g}$
and $E_{2u}$ theories. Our aim is to compare theory and experiment
for $\kappa$ and $\alpha $ in a quantitative fashion. In so doing,
we can also investigate whether fine-tuning is required in either
of these pictures.

In the next section, we give the details of the calculation. The
following section compares the fits for the two models. The final
section discusses how the models can be distinguished on the basis
of pressure experiments and gives a brief summary.

\section{Calculation method}

\subsection{Bands}

The Fermi surface of UPt$_{3}$ is highly complex. There are five
or six sheets, some having a very complicated shape. However, the
effective masses of the sheets have been very accurately
determined for most directions by de Haas-van Alphen experiments
\cite{dhva}. These measurements show that the $\Gamma _{3}$ sheet
contains about 80\% of the density of states at the Fermi energy
$\epsilon _{F}$. This sheet is centered on the $\Gamma $ point. We
approximate it as an ellipsoid with a dispersion relation:

\begin{equation}
\epsilon _{{\bf k}}={\frac{\hbar ^{2}}{2}}\left( {\frac{k_{x}^{2}}{m_{x}}}+{%
\frac{k_{y}^{2}}{m_{y}}}+{\frac{k_{z}^{2}}{m_{z}}}\right)
-\epsilon _{F},
\end{equation}
with $m_{z}=43.2 m_{e}$ and $m_{x}=m_{y}=17.4 m_{e}$. The small
``eggs'' of the $\Gamma _{3}$ band at the zone boundary are
ignored. We have investigated the effect of including other bands
and find that the contributions are small. The anisotropic mass in
the $\Gamma _{3}$ band is very important, however. The results are
quite different for an isotropic Fermi surface.

\subsection{Order Parameters}

The central part of the $\Gamma _{3}$ band does not cross the
Brillouin zone boundary. This means that nodes coming from the
fact that gap functions should be invariant under translation by a
reciprocal lattice vector do not need to be considered. In this
case, to good accuracy, one can use polynomial functions to
describe the gap functions.

Standard calculations for the $E_{1g}$ model use a singlet gap
function
\begin{equation}
\Delta({\bf k}) = \Delta _{0}(T)\,\left[ k_{z} k_{x}+i\,\delta(T)
k_{z} k_{y}\right] ,  \label{eq:del1}
\end{equation}%
while for the $E_{2u}$ model conventional calculations use the
triplet gap function
\begin{eqnarray}
{\bf d}({\bf k}) =\Delta _{0}(T)\hat{z} \left[ k_{z}(k_{x}^{2}-
k_{y}^{2}) + 2 i \delta (T) k_{z} k_{x} k_{y}\right] ,
\label{eq:d1}
\end{eqnarray}
where $k_{\mu }$ is the $\mu$-component of ${\bf k}$ vector on the
Fermi surface and $\hat{z} $ is the direction of the vector ${\bf
d}({\bf k})$ in pseudospin space. Both models involve the idea
that the $A$ phase involves only a single OP component: in the $A$
phase, $\delta (T)=0$. Accordingly, $\delta(T)$ is a
temperature-dependent function that begins to grow continuously at
the $A$-$B$ phase boundary.

A crucial point in the theoretical treatment, however, is that the
expressions in Eqs.~(\ref{eq:del1}) and (\ref{eq:d1}) are {\it
not} the complete OPs for these representations. Yip and Garg
\cite{yip} have shown that the complete gap functions for these
phases are linear combinations of two pairs of functions for
$E_{1g}$ rather than the single pair ($k_{z} k_{x}\,,\,k_{z}
k_{y}$) shown in Eq.~(\ref{eq:del1}). For $E_{2u}$, the gap
function is a linear combination of six pairs rather than the
single pair $\left( \hat{z} k_{z}(k_{x}^{2}-k_{y}^{2})\,,\,
\hat{z} k_{z} k_{x} k_{y}\right)$. Writing the complete gap
functions in the form appropriate for use in both the $A$ and $B$
phases, we have the formulas

\begin{equation}
\Delta ({\bf k})=\Delta_0(T)\sum_{n=1}^{2}g^{(n)}[f_{x}^{(n)}({\bf
k})+i\delta (T)f_{y}^{(n)}({\bf k})] \label{eq:comp1}
\end{equation}
for $E_{1g}$ and

\begin{equation}
{\bf d}({\bf k}) = \Delta _{0}(T) \sum_{n=1}^{6}u^{(n)}[{\bf f} _{x}^{(n)}(%
{\bf k})+i\delta (T){\bf f}_{y}^{(n)}({\bf k})]  \label{eq:compt}
\end{equation}
for $E_{2u}$. The functions $f({\bf k})$and ${\bf f}({\bf k})$ are
defined in Tables~\ref{tab:e1g} and \ref{tab:e2u}, adapted from
Ref.~\cite{yip}. In the $E_{1g}$ representation, the gap function
is a linear combination of the two functions in the first column
of Table~\ref{tab:e1g}. In the $B$ phase, the functions in the
second column also come in, with a relative phase of $i$, as shown
in Eq.~(\ref{eq:comp1}). In the $E_{2u}$ representation, the gap
function is a linear combination of the six functions in the first
column of Table~\ref{tab:e2u}. In the $B$ phase, the functions in
the second column also come in, with a relative phase of $i$, as
shown in Eq.~(\ref{eq:compt}). Note that the resulting gap
function can be multiplied by another completely symmetric
($A_{1g}$) function without changing its symmetry. Nearly all
calculations for $E_{1g}$ have set $g^{(2)}=0,$ while
nearly all calculations for $E_{2u\text{ }}$have set $%
u^{(1)}=u^{(3)}=u^{(4)}=u^{(5)}=u^{(6)}=0$. (For an exception see
Ref.~\cite{hirschfeld}).

\begin{table}[t]
\caption{\label{tab:e1g}Basis functions for the $E_{1g}$
representation.}
\begin{ruledtabular}
\begin{tabular}{lll}
& $f_{x}^{(n)}({\bf k})$ & $f_{y}^{(n)}({\bf k})$ \\ \hline $n=1$
& $k_x k_{z}$ & $k_y k_{z}$ \\  $n=2$ &
$(k_{x}^{5}-10k_{x}^{3}k_{y}^{2}+5k_{x}k_{y}^{4})k_z$ &
$(k_{y}^{5}-10k_{y}^{3}k_{x}^{2}+5k_{y}k_{x}^{4})k_z$ \\
\end{tabular}
\end{ruledtabular}
\end{table}

\begin{table*}
\caption{\label{tab:e2u}Basis functions for the $E_{2u}$
representation.}
\begin{ruledtabular}
\begin{tabular}{lll}
& ${\bf f}_{x}^{(n)}({\bf k})$ & ${\bf f}_{y}^{(n)}({\bf k})$
\\ \hline
$n=1$ & $k_{x}\,\hat{x}-k_{y}\,\hat{y}$ & $k_{y}\,\hat{x}+k_{x}\,%
\hat{y}$ \\
$n=2$ & $(k_{x}^{2}-k_{y}^{2})k_{z}\hat{\,z}$ & $2\,k_{x}k_{y}k_{z}\,%
\hat{z}$ \\
$n=3$ & $(k_{x}^{3}-3k_{x}k_{y}^{2})\,\hat{x}%
-(k_{y}^{3}-3k_{y}k_{x}^{2})\,\hat{y}$ & $(k_{x}^{3}-3k_{x}k_{y}^{2})\,%
\hat{y}+(k_{y}^{3}-3k_{y}k_{x}^{2})\,\hat{x}$ \\
$n=4$ & $(k_{x}^{3}-3k_{x}k_{y}^{2})\,\hat{x}%
+(k_{y}^{3}-3k_{y}k_{x}^{2})\,\hat{y}$ & $(k_{x}^{3}-3k_{x}k_{y}^{2})\,%
\hat{y}-(k_{y}^{3}-3k_{y}k_{x}^{2})\,\hat{x}$ \\
$n=5$ & $(k_{x}^{4}-6k_{x}^{2}k_{y}^{2}+k_{y}^{4})k_{z}\,\hat{\,z}$ & $%
(4k_{x}^{3}k_{y}-4k_{x}k_{y}^{3})k_{z}\,\hat{\,z}$ \\ $n=6$ &
$(k_{x}^{5}-10k_{x}^{3}k_{y}^{2}+5k_{x}k_{y}^{4})\hat{x}
+(k_{y}^{5}-10k_{y}^{3}k_{x}^{2}+5k_{y}k_{x}^{4})\hat{y}$ &
$(k_{x}^{5}-10k_{x}^{3}k_{y}^{2}+5k_{x}k_{y}^{4})\hat{y}
-(k_{y}^{5}-10k_{y}^{3}k_{x}^{2}+5k_{y}k_{x}^{4})\hat{x}$
\\
\end{tabular}
\end{ruledtabular}
\end{table*}

It is very important to understand what this special choice of $g^{(n)}$ or $%
u^{(n)}$ implies. There are two levels of doing phenomenology for
the OP. One is choosing a representation and working out the
consequences. In this case the coefficients $g^{(n)}$ and
$u^{(n)}$ must be treated as parameters. They are {\it continuous}
functions of all the many parameters that are in the microscopic
Hamiltonian, as well as external variables such as pressure. Thus
choosing a representation means choosing a {\it set} of
Hamiltonians of finite measure, i.e., to selecting an alternative
that has a finite {\it a priori} probability. The other (more
commonly used) level of phenomenology is to choose a specific
basis function from Table~\ref{tab:e1g} or Table~\ref{tab:e2u} and
work out the consequences. This means setting all but one of the
$g^{(n)}$ or $u^{(n)}$ equal to zero. It also means choosing a set
of Hamiltonians of zero measure, which has zero {\it a priori
}probability. In other words, choosing a specific basis function
amounts to fine tuning. None of the $g^{(n)}$ or $u^{(n)}$ can
vanish by symmetry, since all symmetries of the system have
already been taken into account in the group-theoretical
decomposition.

This does not necessarily invalidate previous work that has been
done with a specific choice of basis function. If the comparison
to experimental data is not very sensitive to variations in
$g^{(n)}$ or $u^{(n)},$ then the fine tuning may be abandoned
without harming agreement of theory and observation. A key goal of
this paper is to investigate this sensitivity for the $E_{1g}$ and
$E_{2u}$ theories.

\subsection{Transport Coefficients}

Ultrasonic attenuation experiments in UPt$_{3}$ have been done in
the hydrodynamic regime where the sound frequency $\omega $
satisfies $\omega<1/\tau _{s}$, with $\tau _{s}$ the electronic
relaxation time. The appropriate formula for sound propagation
along the direction $\hat{q}$ with polarization $\hat{\varepsilon
}\perp \hat{q}$ is:

\begin{widetext}
\begin{eqnarray}
\alpha _{\hat{\bf q}\hat{\bf \varepsilon}}(T)\propto
\int_{0}^{\infty }d\omega \left[ -{\frac{\partial F(\omega
)}{\partial \omega }}\right] \tau _{s}(\omega ,T)\left\langle {\rm
Re}\left( {\frac{\sqrt{\omega ^{2}-|\Delta _{{\bf
k}}|^{2}}}{\omega }}(\hat{{\bf q}}\cdot \hat{{\bf
k}})^{2}(\hat{{\bf \varepsilon }}\cdot \hat{{\bf k}})^{2}\right)
\right\rangle . \label{eq:alpha}
\end{eqnarray}
\end{widetext}
Here $F$ is the Fermi function and the angle brackets indicate a
Fermi surface average. $|\Delta _{{\bf k}}|^{2}\rightarrow |{\bf
d}_{{\bf k}}|^{2}$ in the triplet case. The thermal conductivity
in the $i$th direction when a temperature gradient is imposed
along the $i$th direction is computed using:

\begin{widetext}
\begin{eqnarray}
\kappa_{ii}(T)\propto {1\over T}\int_0^\infty d\omega
~\omega^2\left[-{\partial F(\omega)\over \partial \omega}\right]
\tau_s(\omega,T)\left\langle{\rm Re}
\left({\sqrt{\omega^2-|\Delta_{\bf k}|^2}\over \omega}
(\hat{i}\cdot\hat{\bf k})^2 \right)\right\rangle. \label{eq:kappa}
\end{eqnarray}
\end{widetext}
The prefactors are fitted to the values of $\alpha _{\hat{\bf
q}\hat{\bf \varepsilon}}$ and $\kappa _{ii}$ at $T_{c+}$.

Our approach does not calculate the density of states
self-consistently. This means that it is not very accurate at low
temperatures, $T < 0.2 T_{c+},$ but in this paper we are chiefly
concerned with the behavior closer to $T_{c+}$.

The temperature dependence of $\Delta _{0}(T)\,$ is not accurately
known. \ However, one may deduce the power law at low temperature
from the pattern of nodes, while near $T_{c}$ we can use
Ginzburg-Landau theory for the same purpose. For computational
convenience, we take $\delta (T)=[1-(T/T_{c-})^{6}]^{1/2}$ and
$\Delta _{0}(T)=\Delta _{0}(T=0)\left[ 1-(T/T_{c+})^{x}\right]
^{1/2}.$ $x=2$ and $x=3$ for point and line nodes, respectively.
We have verified that these functions are close to the
Ginzburg-Landau result near $T_{c+}$ and they have the appropriate
behavior at low temperature as well. In addition, this leads to a
fit for the temperature dependence with only the single parameter
$\Delta _{0}(T=0)$. Finally, we follow Graf {\it et al.}
\cite{sauls2} in taking the empirical form for the relaxation
time: $\tau _{s}\simeq 0.01\pi T_{c+}(1+T^{2}/T_{c+}^{2})$.

\subsection{Domains}

In the A phase there is an additional complication. $\delta (T)=0$
so that, as stated above, only the $f_{x}({\bf k})$ (or ${\bf
f}_{x}({\bf k})$) functions are used. However, the theory requires
that the superconducting OP is oriented by its coupling to the
staggered antiferromagnetic moment ${\bf M}_{S}.$ There are three
domains allowed by symmetry for ${\bf M}_{S}$, and neutron
scattering experiments show that they are equally populated. The
pattern of nodes rotates by $\pi /6$ around the $c$-axis every
time the wave passes through a domain wall. The gap function
becomes, for example, $f_{x}(\mathcal{R}\,{\bf k})$ where
$\mathcal{R}\,{\bf k}$ is the rotated vector. We shall take a
coordinate system which is fixed in the laboratory: the
propagation vector or polarization vectors refer to this frame.

The observed $\alpha$ must result from an average over domains. In
such a situation, response functions are usually calculated by
effective medium theory, which may take various forms, depending
on the approximations made \cite{choy}. However, the present case
is simplified by the fact that $\alpha$ is the relatively small
imaginary part of a large complex quantity, the sound propagation
vector. For such a perturbative quantity, all averaging methods
lead to the result that the observed global $\alpha$ should be the
mean of the three local values.

%%%%%%%%%%%%%%%%%%%%
\begin{figure}[t]
\includegraphics[scale=0.68]{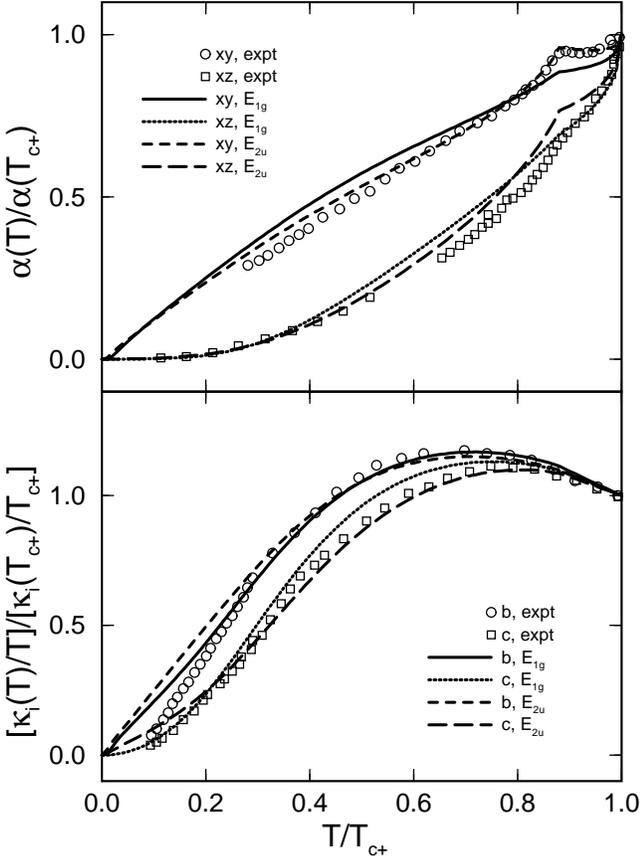}
\caption{\label{fig1} Best fits of $E_{1g}$ and $E_{2u}$ models
against the data \protect\cite{ellman96,lussier96} for both
transverse sound attenuation (top frame) and thermal conductivity
(bottom frame). The fitting parameters are given in the text.}
\end{figure}
%%%%%%%%%%%%%%%%%%%%

\section{Results}

\subsection{Best fits}

We begin the comparison of theory and experiment in this
subsection by adjusting all parameters in Eqs.~(\ref{eq:comp1})
and (\ref{eq:compt}) to their optimum values for both
representations, varying the mass anisotropy $m_z/m_x$ and
choosing the most favorable domain of the $A$ phase. Fairly
impressive fits of data to experiment for both theories may be
obtained in this way.

%%%%%%%%%%%%%%%%%%%%%%%%%
\begin{figure}[t]
\includegraphics[scale=0.68]{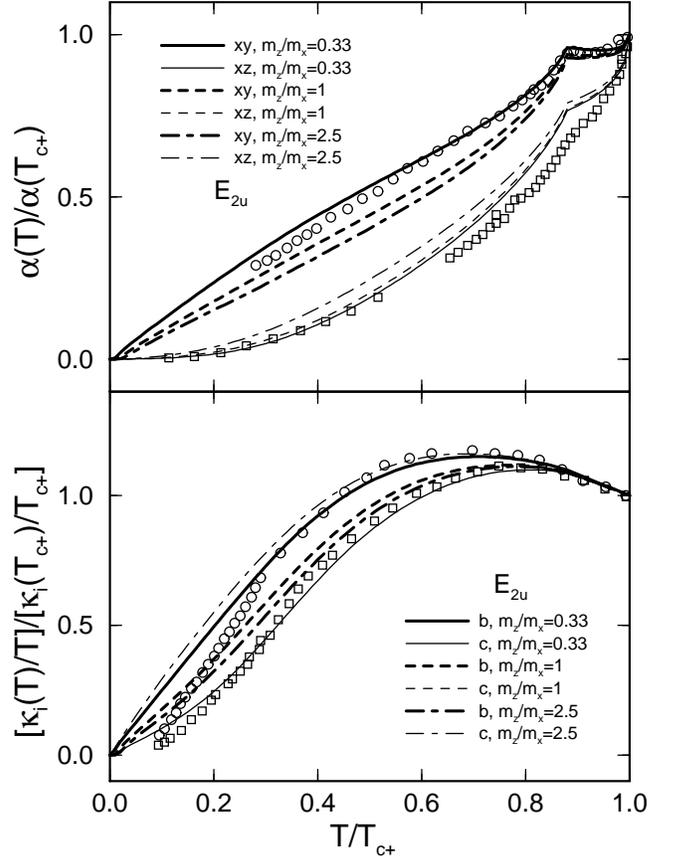}
\caption{\label{fig2} The figure shows the sensitivity of the fits
to the change of the anisotropic mass ratio. In the bottom frame,
the two curves ($b$ and $c$) are indistinguishable for the
$m_z/m_x=1$ case.}
\end{figure}
%%%%%%%%%%%%%%%%%%%%%%%%%

The best fits for $E_{1g}$ and $E_{2u}$ models for the ultrasound
and the thermal conductivity are plotted against the data in
Fig.~\ref{fig1}. The curves for $\kappa$ provide reasonable fits
over the whole range of temperature. In particular, the overall
{\em anisotropy} is well fit. Some discrepancies may be noted at
sufficiently low temperature. The fits for $\alpha$ are also
reasonable except that some minor discrepancies may be noted near
$T_{c}$ in the $A$ phase. The best-fit parameters for $E_{1g}$
are: $\Delta(0)=3T_{c+}$, $x=2$, and $g^{(1)}=1.0$ and
$g^{(2)}=0.2$, while the best-fit parameters for $E_{2u}$ are:
$\Delta(0)=4.5T_{c+}$, $x=2$, and all the coefficients $u^{(i)}$
vanish except $u^{(2)}=1.0$.

There exists a major difference between the best fits of $E_{1g}$
and $E_{2u}$ models when anisotropic masses are considered. For
$E_{1g}$, the anisotropic mass ratio is fixed as $m_z/m_x=m_z/m_y
= 2.5$.  This agrees with the measured value \cite{dhva} for the
dominant $\Gamma_3$ band. For $E_{2u}$, in contrast, the best fits
are acquired at $m_z/m_x=m_z/m_y=0.33$, in very poor agreement
with the measured value.

\subsection{Anisotropic mass}

In order to give some idea of the sensitivity of the fits to  the
anisotropic mass ratio, we calculate $\alpha$ and $\kappa$ for
$E_{2u}$ model using the same best-fit parameters except for the
anisotropic mass ratio $m_z/m_x$. As seen in Fig.~ \ref{fig2}, as
the ratio is increased from 0.33 to its physical value of 2.5, the
polarization anisotropy in $\alpha$ decreases sharply, while the
anisotropy in $\kappa$ actually reverses. This indicates that the
fit for the $E_{2u}$ model is somewhat questionable.

Our fitting procedure differs from that of Graf {\it et al.}
\cite{graf96,sauls2}. These authors used revised order parameters

\begin{equation}
\Delta({\bf k}) = \Delta _{0}(T) \left[ {k}_{z}{k}_{x}+i\,\delta
(T)\,{k}_{z}{k}_{y}\right](1+a_2 {k}_{z}^2+a_4 {k}_{z}^4)
\end{equation}
for $E_{1g}$ and

\begin{equation}
{\bf d} ({\bf k}) =\Delta_0(T) \hat{z}
[{k}_{z}({k}_{x}^{2}-{k}_{y}^{2})+ 2 i \delta(T){k}_{z} {k}_{x}
{k}_{y}] (1+a_2 {k}_{z}^2+a_4 {k}_{z}^4)
\end{equation}
for $E_{2u}$, where the coefficients $a_2$ and $a_4$ are
adjustable fitting parameters. Since the additional factor is
completely symmetric, this does not change the representation.  As
opposed to using the standard basis functions ($a_2=a_4=0$), it is
evident that large $a_2$ and $a_4$ will correspond to more
electronic excitations around the line of nodes at $k_z=0$. This
in turn will enhance the in-plane thermal conductivity and sound
attenuation relative to the out-of-plane ones. Roughly speaking,
the larger $a_2$ or $a_4$ case for the revised model is physically
analogous to having $m_x,m_y>m_z$ (and hence larger in-plane
density of states) in the standard model.  This change in the gap
function amounts to multiplying the relative contribution of the
in-plane excitations by a factor of 12 relative to what one would
expect from the measured ratio of effective masses.

\subsection{Domain-averaged fits}

In order to fit the data in the $A$ phase, it is necessary to
choose the most favorable domain. This was also found by Graf {\it
et al.} \cite{sauls2}. As mentioned above, this is not consistent
with the assumptions of the theory. The idea that the order
parameter is oriented by ${\bf M}_{S}$ is a central assumption for
both pictures, since it accounts for the fact that $H_{c2}$ is
isotropic in the basal plane. Thus domain averaging is not
optional: it is necessary for the consistency of the theory.

%%%%%%%%%%%%%%%%%%%%%%%%%
\begin{figure}[t]
\includegraphics[scale=0.68]{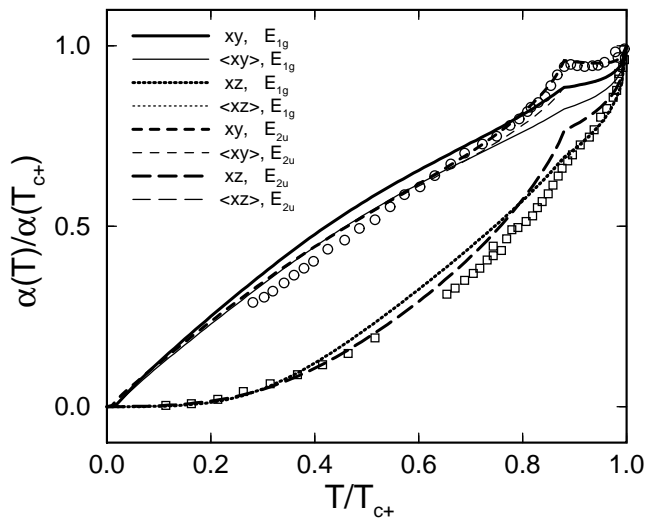}
\caption{\label{fig3} The best fits of ultrasound attenuation data
taking into account three equal-populated domain average. The best
fits for one single domain (see also Fig.~\ref{fig1}) are included
for comparison. The curves for $xz$ and $\langle xz\rangle$ are
indistinguishable.}
\end{figure}
%%%%%%%%%%%%%%%%%%%%%%%%%

If the absorption is properly averaged over all three domains,
then we obtain the best fit for the two theories shown in
Fig.~\ref{fig3}. In the $A$ phase, the fits are substantially
worse than those one gets by choosing the best domain, as was done
for Fig.~\ref{fig1} and in Ref.~\cite{sauls2}.

The reason for the difference in the quality of fit in the $A$
phase is mainly due to the node patterns. In the $B$ phase, for
both $E_{1g}$ and $E_{2u}$ models, there are lines of nodes along
the equator and point nodes at the poles of the Fermi surface. In
this phase, the nodal pattern is symmetric about the $c$-axis and
thus the real nodal structures are virtually the same in the three
different domains. In the $A$ phase, in contrast, the two point
nodes at the pole have connected to form line nodes in the $k_x=0$
plane for $E_{1g}$ and the $k_x=\pm k_y$ planes for $E_{2u}$ model
(in the domain defined by the basis functions of
Tables~\ref{tab:e1g} and \ref{tab:e2u}). In this case, the nodal
structure is asymmetric about the $c$-axis. Depending on the
orientation of the node pattern relative to the direction of
polarization/propagation, the three domains contribute
differently. As a matter of fact, as compared to an isotropic
state, transverse sound with $\hat{{\bf q}} \parallel \hat{x}$ and
$\hat{{\bf \epsilon}}\parallel\hat{y}$ (denoted as $xy$) will
cause fewer excitations in the $\pi/6$ or $\pi/3$ rotated domains.
It is this that causes the $\langle xy\rangle$ (the angle brackets
indicate a three equal-populated domains average) attenuation to
drop significantly in the $A$ phase relative to the $xy$
attenuation.  In contrast, the best fits for the $xz$ and $\langle
xz\rangle$ attenuations are almost indistinguishable. These
considerations apply equally to the $E_{1g}$ and $E_{2u}$ models.

\subsection{Fine tuning}

An important criterion for the acceptance of a theory is its
robustness. If small changes in the parameters would spoil the
agreement with the data, the theory is not acceptable since it
lacks explanatory power. Accordingly, we investigate the
sensitivity of the OP's transport characteristics to changes in
its shape. We shall focus on the $B$ phase for reasons that will
become clear below.

\begin{figure}[t]
\includegraphics[scale=0.65]{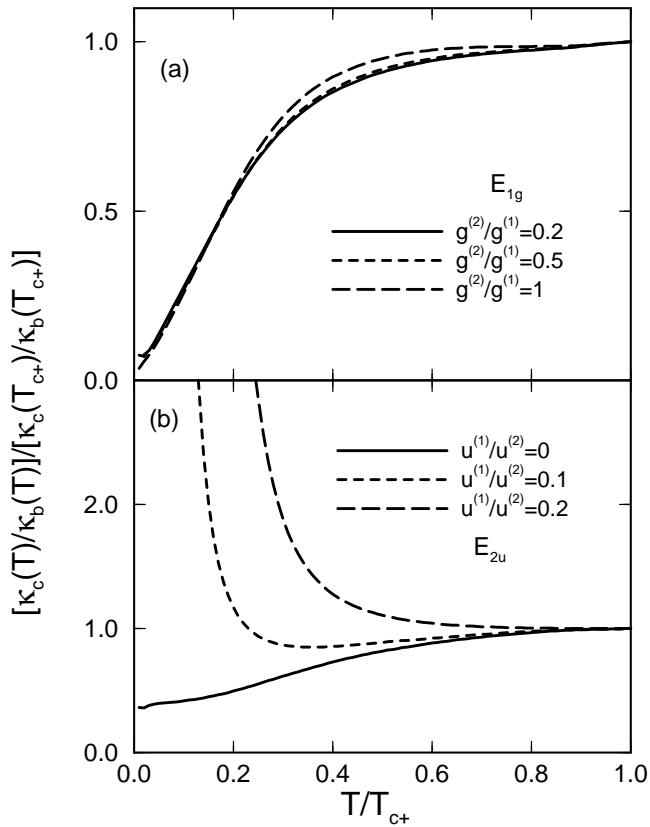}
\caption{\label{fig4} (a) The ratio of $\kappa_c/\kappa_b$ for
different ratios of $g^{(2)}/g^{(1)}$ for the $E_{1g}$ model. (b)
The ratio of $\kappa_c/\kappa_b$ for different ratios of
$u^{(1)}/u^{(2)}$ for the $E_{2u}$ model.}
\end{figure}

For $E_{1g},$ varying the shape is relatively straightforward.
There are only two parameters $g^{(1)}$ and $g^{(2)}$. In the $B$
phase with $g^{(2)}=0 $ there are lines of nodes along the equator
and point nodes at the poles. Changing the ratio of $g^{(1)}$ and
$g^{(2)}$ does not affect this nodal pattern. A larger value of
$g^{(2)}$ merely gives additional modulation to the order
parameter in the azimuthal direction. We show some representative
curves for $\kappa_c/\kappa_b$ for different values of the ratio
in Fig.~\ref{fig4}(a). There is very little change in the goodness
of fit as the ratio is varied. Defining a chi-square parameter for
the fits as a function of $g^{(2)}/g^{(1)}$, we see from its
approximate constancy that the $E_{1g}$ theory is robust for this
particular quantity.

%%%%%%%%%%%%%%%%%%%%%%%%%%%
%%%%%%%%%%%%%%%%%%%%%%%%%%%

The $E_{2u}$ order parameter behaves very differently. For
starters, the shape space $u^{(1)},...,u^{(6)}$ is
six-dimensional. This precludes an extensive search. However,
mapping of a two-dimensional subspace will suffice to make the
point about parameter sensitivity. The main difference between
$E_{2u}$ and $E_{1g}$ is clear from the nodal pattern. Taking into
account $u^{(1)}$ in addition to the usual $u^{(2)}$, when
$u^{(1)}=0$ there is a line of nodes at the equator and point
nodes at the poles in the $B$ phase. When $u^{(1)}$ is finite,
however, the line of nodes disappears. This is a consequence of
Blount's theorem \cite{blount85}, which states that triplet states
in a spin-orbit coupled system do not have lines of nodes. In
Fig.~\ref{fig4}(b), we show some representative curves for $\kappa
_{c}/\kappa _{b}$ for different values of the ratio
$u^{(1)}/u^{(2)}$. The curves are very sensitive to the ratio, in
particular at low temperatures. The chi-square parameter varies
rapidly as a function of $u^{(1)}/u^{(2)}$. We conclude that the
$E_{2u}$ theory is not robust. Its virtues in describing the data
arise from a fine-tuning of the parameters.

\section{Conclusion}

Order parameters belonging to the two-dimensional representations
$E_{1g}$ and $E_{2u}$ remain as the main candidates for the
superconducting state of UPt$_{3}$. A difficulty that remains for
both these theories is the explanation of the ultrasonic
absorption in the $A$ phase. The theories require that the
observed absorption be the result of an average over three domains
in which the order parameter has three orientations. This
averaging is deleterious to the agreement between theory and
experiment. This may indicate that there is something wrong with
the idea that the magnetization orients the superconductivity, or
it may indicate that there is an additional scattering mechanism,
such as scattering from domain walls, which partially cancels the
drop in absorption due to gap formation.

The $E_{2u}$ theory has an additional severe problem at low
temperatures. A line of nodes at the equator of the Fermi surface
is required in order to give agreement with experiment for the
thermal conductivity. In the six-dimensional parameter space of
this theory, this occurs only in a two-dimensional subspace. The
theory therefore requires fine-tuning to explain the data: all the
parameters in the microscopic interaction must conspire to give
the desired result. It has been suggested that strong spin-orbit
coupling can lock the ${\bf d}$-vector along the $z$-direction.
However, ${\bf d}$ does not have the interpretation of zero spin
projection, as it does in $^{3}$He. Taking into account the
spin-orbit coupling reduces the order parameter space to six
dimensions for triplet $E_{2}$, not one or two.

Since the shape parameters $u^{(n)}$ depend on all the details of
the microscopic Hamiltonian, the low temperature power laws that
depend on having a line of nodes are unstable. By contrast, the
line of nodes at the equator in the $E_{1g}$ theory is forced by
symmetry.  Application of {\em pressure}, for example would lift
these nodes in an $E_{2u}$ gap. The pressure scale that governs
superconducting phenomena may be estimated as
$T_{c+}(dT_{c+}/dP)^{-1} \sim 1 {\rm kbar}$.  One would expect
very strong pressure dependence of the thermal conductivity on
this scale at low temperatures for $E_{2u}$ but not for $E_{1g}$.
This could serve to further distinguish the two theories.

The data presently give preference to the $E_{1g}$ theory since it
does not require fine tuning to explain the data.  The discrepancy
of theory and experiment for ultrasonic absorption in the $A$
phase has not been resolved, and may require new ideas.

\begin{acknowledgments}
Financial support from NSC of Taiwan (Grant No. 90-2112-M-003-018)
and the NSF under the Materials Theory program, Grant No.
DMR-0081039 is acknowledged.
\end{acknowledgments}

%\bibliography{upt3}%

\end{document}